\documentclass{article}

\author{N. Redington and M.A.K. Lodhi
\thanks{Corresponding author: a.lodhi@ttu.edu}
\\ Department of Physics, \\Texas Tech University,
\\MS 1051, Lubbock, TX 79409 USA}
\title{A Simple Five-Dimensional Wave Equation for a Dirac Particle}
\date{}

\begin{document}

\maketitle

A first-order relativistic wave equation  
is constructed in five dimensions. 
Its solutions  
are eight-component spinors, which are  
interpreted as  single-particle fermion  wave functions  
in four-dimensional spacetime.  
Use of a ``cylinder condition'' (the removal of explicit 
dependence on the fifth co\"{o}rdinate) 
reduces each eight-component  solution  
to a pair of  degenerate four-component spinors  
obeying the Dirac equation. 
This five-dimensional method  
is used to obtain solutions  
for  a free particle and for a particle 
moving in the Coulomb potential. It is shown that, under the cylinder
condition, the results are the same as those from the Dirac equation. 
Without the cylinder condition, on the other hand, 
the equation predicts some interesting new phenomena. It implies 
the existence of a scalar potential, and for zero-mass particles
it leads to a four-dimensional fermionic equation analogous to
Maxwell's equation with sources.

\bigskip
PACS Number: 03.65.Pm

\newpage

\section{Introduction}

The Dirac equation correctly describes the behaviour of a single 
relativistic fermion. However, it is hard to solve and also hard
to visualise. In this paper, we study a mathematically somewhat
simpler (but five-dimensional) first-order linear 
equation, and show that it is entirely
equivalent to Dirac's. 

Several previous  authors, beginning with Dirac
himself [1],  have considered five-dimensional
generalisations of the one-body Dirac equation. 
In most cases (e.g. [2-6]), these authors have used   a 
spacelike fifth dimension.  This  choice of metric
has been the usual one for higher dimensional physical 
models since the time of   Kaluza, in whose  unified 
theory of gravity and electromagnetism it is in fact a
\begin{em} necessary\end{em} assumption
if Maxwell's equations are to have the correct sign [7].  

If we avoid Kaluza's interpretation of  the fifth component of momentum
as charge, we are free to use a second time-dimension rather than a fourth space-dimension.
This has been done in classical general relativity by numerous authors (e.g. Kocinski [8]).
In the ``induced matter'' version of Kaluza theory developed by Wesson [9-12], 
Ponce de Leon [13, 14], and others, ordinary massive particles in four-dimensional
space-time 
are treated as massless neutrinos in a curved, non-compact five-dimensional space
equivalent to the ``bulk'' space of membrane theory. Induced matter has
been discussed in terms of both one-time and two-time metrics. 

Also relevant is the work of  Bars and his 
collaborators [15-17], who have proposed  that the Standard Model is simply a gauge-fixed 
form of some two-time (and four-space) theory. Their six-dimensional approach, although
developed independently, 
in some ways closely parallels an earlier one
originating with Dirac [18-21]. For Dirac, the goal was to 
explain four-dimensional physics in terms of conformal geometry; for Bars,
it is to reveal different four-dimensional dynamical systems
as ``holographic views'' of the same six-dimensional system differing in 
gauge choice. 

Like Dirac and Bars, we  choose the new dimension to be timelike, but  retain only
the conventional three space dimensions, thus abandoning the conformal and 
holographic interpretations of the higher-dimensional manifold. In our model,
the wave function of  a single fermion is represented by a spinor with eight
complex components. A first order linear wave equation in five-dimensional 
spacetime governs the behaviour of the wave function. We then introduce a
constraint in the spirit of Kaluza's ``cylinder condition'' which prevents the fifth
dimension from appearing explicitly in the final results. With this constraint in
place, the eight-component spinor wave function reduces to two coupled
four-component spinors, both of which obey the ordinary four-dimensional 
Dirac equation. When the cylinder condition is \begin{em}
not\end{em} imposed, the five-dimensional wave equation may be viewed 
as a pair of coupled four-dimensional equations, with possible new physical
effects implied by the coupling. For example, we will find that 
when one of the two coupled spinors is held constant over
a four-dimensional region, the other behaves like a conventional
Dirac wave function in the presence of a scalar potential.

Eight-component spinors, or at least pairs of four-component spinors, 
and double copies of the Dirac equation, have occasionally 
been used (e.g. by  Joyce [22]) even without any reference to 
higher dimensional space.  The earliest example seems to be the  work of Lanczos [23, 24], who  
in 1929 proposed two coupled four-dimensional
quaternionic ``wave equations'' which he 
showed were equivalent to two independent copies of the Dirac equation.
We will find that the somewhat cumbersome equations discovered by Lanczos are
a limiting case of our simpler five-dimensional equation.  

The paper is organised as follows:

In Section II, we review the standard geometric algebra approach to the
four-dimensional Dirac equation.
In Section III, we  extend this approach to five dimensions, presenting our 
new wave equation and showing that it 
reduces to the standard four-dimensional free particle Dirac equation when
a ``cylinder condition'' is used to eliminate explicit dependence on the 
second time dimension.  In Section IV,
we find plane-wave solutions of this equation which correspond to
Dirac free particles.  
In Section V, as an example of our approach, we
solve the new wave equation with the cylinder condition for the
case of a Coulomb potential, obtaining the standard hydrogen atom spectrum.
Finally, in Section VI, we relax the cylinder condition and find that 
the five-dimensional equation predicts several interesting new effects,
notably the existence of a scalar field.

\section{Geometric Approach to the Dirac Equation}

Throughout this paper, we use 
Clifford (geometric) algebra techniques to handle vectors and spinors; 
geometric algebra and its applications to physics are exhaustively reviewed in [25-30].
The Clifford algebra describing a flat spacetime with $m$ positive-norm
and $n$ negative-norm unit vectors  is  called $C\ell (m,n)$;  for example, the Clifford
algebra associated with Minkowski space is   $C\ell (3,1)$. The
appropriate Clifford algebra for our five-dimensional case is $C\ell (3,2)$.

Note that we are  using the ``$+++-$'' convention for the metric of flat spacetime: 

\begin{equation}
dx^{A} dx_{A} = dx^{\mu} dx_{\mu} - (dx^{4})^{2} = dx^{i} dx_{i} - (dx^{0})^{2} - (dx^{4})^{2} =   ds^{2}
\end{equation}
The 
point is significant, because  $C\ell(n,m)$ and $C\ell(m,n)$ are not
generally isomorphic. 
(In Eq. (1)  Latin lower-case indices run over the space co\"{o}rdinates from 1 to 3.
Greek lower-case indices run from
0 to 3, where $x^0$ is the ordinary time. Latin upper-case indices run 
from 0 to 4, where $x^4$ is the new, ``extraordinary'' time dimension.
We employ ``natural'' units in which $c$ and $\hbar$ are $1$.) 

In four dimensions, 
the traditional covariant matrix formulation of the Dirac equation with the $+++-$ metric [31] 
\begin{equation}
\gamma^{\mu} \partial_{\mu} |\Psi> = - m |\Psi> 
\end{equation}
(where the $\{\gamma^{\mu}\}$ are the Dirac gamma matrices, $|\Psi>$ is the
state vector, and $m$ is mass)
can be translated into a matrix-free expression in the language of geometric algebra 
by a two-step process. First one places the real and imaginary parts of the
four components of the column-spinor $|\Psi>$ in one-to-one
correspondence with the eight \begin{em}real\end{em} components of $\Psi$, a general 
even-grade multivector of $C\ell(3,1)$. Then one directly computes the effect of   
matrix multiplication by
$\gamma^{\mu}$ on $|\Psi>$ in some given representation and seeks a Clifford algebra
operator having the same effect on the multivector $\Psi$. In the Pauli-Dirac
representation of the gamma matrices, it can be readily verified that the effect
of the $\gamma^{\mu}$ on a column matrix are the same as that of the operator
which first left-multiplies the multivector $\Psi$ 
by $e^{\mu}$ and then right-multiplies it by $e^{0} e_{1} e_{2}$. 
(Here the $\{e_{\mu}\}$ are unit vectors in Minkowski 
spacetime.) Replacing the gamma matrices in the conventional Dirac equation 
by their Clifford algebra 
equivalents leads immediately to the so-called ``Hestenes form'' of the Dirac equation [30]:  

\begin{equation}
e_{\mu} \partial ^ {\mu}  \Psi = m \Psi e_{0} e_{1} e_{2} 
\end{equation}

Despite its appearance, this representation of the Dirac equation is covariant and 
completely interchangeable with the matrix version, as discussed 
in [32-36].  

(We note in passing that, as shown by Lounesto [34], Eq. (3) also holds in the $---+$ metric,
the lack of isomorphism between $C\ell(3,1)$ and $C\ell(1,3)$ notwithstanding.
The reason for this is that the appropriate Clifford equivalent of 
$\gamma^{\mu} |\Psi>$ in this metric turns out to be $e^{\mu} \Psi e^{0}$. Because
the Dirac equation in the $---+$ metric has a factor of $i$ in the right-hand side
of Eq. (2) and $i |\Psi>$ corresponds to $\Psi e_{1} e_{2}$ in either metric [26], 
the final result is once again Eq. (3). Nearly all authors who have previously studied 
Eq. (3) have worked in the $---+$ metric.)

Eq. (3) involves both
right- and left-multiplication, somewhat complicating its solution. 
The main advantage of   Eq. (3)
over Eq. (2) is not so much computational as conceptual. In particular, Eq. (3)
highlights the often-overlooked correlation between the dimensionality of 
spacetime and the number of components possessed by a spinor. 

That such a correlation exists is suggested  by the fact that Dirac spinors, appropriate 
to four-dimensional spacetime, have four complex components; non-relativistic Pauli spinors,
appropriate to three-dimensional space, have two; and the simple Schr\"{o}dinger
wave function without spin has only one.  In the matrix approach this trend is without
obvious explanation, but it
follows naturally from the geometric algebra approach to quantum theory.   
In the general $(m+n)$ dimensional case,  a complex column-vector with 
$2^{m+n-2}$ components can always be placed in one-to-one correspondence with 
an even real element of  $C \ell (m,n)$. Thus, if  we assume that a spinor is 
by definition an even element of the algebra,  the dimensionality of its column-vector
representation follows automatically [37].
 
It follows that  in five dimensions
we expect to write the wave function as an eight-component 
complex-valued spinor (or, equivalently, as a sixteen-component real-valued even 
element of the algebra).  If this wave function is to represent a \begin{em}single\end{em}
particle, we must either introduce a restriction which eliminates half of the components
or else accept the existence of  two distinct classes of  fermions. We will return to this issue later.

The appropriate five-dimensional generalisation of 
the various terms in Eq. (3) depends on the value of  $k$  in $C \ell (5-k,k)$,
that is, on the number of timelike dimensions.  In  Section III  below we will see  that for
$k=2$ it is 
possible to  find a wave equation which, at least in the free particle case, 
involves only left-multiplication, thus avoiding one of the disadvantages of
Eq. (3).

\section{Five-Dimensional Wave Equation for a Free Particle}

Let $E$ be the pseudoscalar ``volume'' element of $C\ell(3,2)$:

\begin{equation} 
E=e_{0} \wedge e_{1} \wedge e_{2} \wedge e_{3} \wedge e_{4}
\end{equation}
where $\{e_{A}\}$ are unit vectors. Note that $E^{2} = 1$. Because
$E$ is pseudoscalar, it commutes with every even-grade element
of the algebra; because the dimension of the spacetime happens to
be odd, it also commutes with every odd-grade element. This contrasts
sharply with the four-dimensional case, in which there exists no
non-scalar element commuting with every multivector. 

Consider the first-order wave equation:

\begin{equation}
e_{A} \partial^{A} \Phi = - E m \Phi
\end{equation} 
where $\Phi$ is an even element of $C\ell(3,2)$, corresponding to an
\begin{em}eight \end{em} component complex spinor in the more usual matrix notation.
The relativistic invariance of Equation (5) is evident. Only left-multiplication is
used, although because $E$ and $\Phi$ commute, we could also have written the
right side of (5) as $- m \Phi E$. Eq. (5) is our proposed five-dimensional field 
equation for the free particle, which we wish to prove equivalent to Dirac's.

Note that if we left-multiply both sides of Eq. (5) by $e_{B} \partial^{B}$
we obtain the five-dimensional Klein-Gordon equation. Here we see the significance
of using two times instead of  four space-co\"{o}rdinates: if we had chosen $C\ell(4,1)$,
$E^{2}$ would be $-1$ and  the Klein-Gordon equation would have the wrong sign.

We will now demand that 

\begin{equation}
\partial^{4} \Phi = 0
\end{equation} 
that is, that there be no explicit dependence of $\Phi$ on the new ``extraordinary''
time co\"{o}rdinate. (This requirement is of course reminiscent of the ``cylinder condition''
of Kaluza's unified theory [7],  and we will call it by that name for convenience; 
note however that Kaluza's fifth dimension was 
space-like.) We claim that, with this restriction,
the pseudoscalar field equation decomposes into
two independent copies of the usual Dirac equation.

To show this, we will project Eq. (5) onto ordinary Minkowki space.
Consider the operation $(\cdot)_{\pm}$, defined by

\begin{equation}
\Phi_{\pm} \equiv \frac{1}{2} (\Phi \pm e^{4}\Phi e_{4})
\end{equation}
This operation separates those Clifford blades in $\Phi$ which contain
factors of $e^{4}$ from those which do not. Similar
``rejection/projection'' operators are used routinely in Clifford
algebra, and their properties are well-known [30]. In our case, since
$\Phi$ is an even-grade element of $C\ell(3,2)$, 
it is easy to show that
$\Phi_{+}$ contains \begin{em}no\end{em} terms of the form $e_{\mu} \wedge e_{4}$,
while $\Phi_{-}$ contains \begin{em}only\end{em} such terms -- hence the names
``rejection'' and ``projection'', meaning of $\Phi$ onto $e_{4}$.

From the definition (7) and the fact that 

\begin{equation}
e^{4} \wedge e_{\mu} = - e_{\mu} \wedge e^{4} 
\end{equation} 
we see that 

\begin{equation}
(e_{\mu}\Phi)_{\pm} = e_{\mu} \Phi_{\mp}  
\end{equation}
Applying the $(\cdot)_{\pm}$ operator to Eq. (5) and using Eq. (9), we find
(after recalling that the pseudoscalar $E$ commutes with every spinor)

\begin{equation}
 e_{4} \partial ^{4} \Phi_{\pm} + e_{\mu}\partial ^ {\mu} \Phi_{\mp} = - m\Phi_{\pm}  E 
\end{equation}
which can be viewed as a pair of coupled equations relating $\Phi_{+}$
to $\Phi_{-}$. 

In fact, if we were to drop the first term (as we eventually will when we invoke the cylinder condition)
and replace $E$ by $\pm i$
we would have the two coupled field equations
suggested  by Lanczos [23] in 1929. Lanczos of course made no reference to the fifth dimension,
and therefore had no simple underlying equation like Eq. (5). He interpreted his two fields, equivalent to our $\Phi_{+}$ and $\Phi_{-}$, as independent 
four-component spinors.    

We now define 
\begin{equation}
\Xi  \equiv  \Phi(1-e_{3}e_{4}) 
\end{equation} 
It will prove significant that $\frac{1}{2}(1-e_{3}e_{4})$ is idempotent;  Lanczos 
also used idempotent multipliers in the process of  going from his field equations to the 
conventional Dirac equation.

Eq. (11) can be rewritten as 

\begin{equation}
\Xi_{\pm} \equiv \Phi_{\pm} - \Phi_{\mp} e_{3}e_{4}
\end{equation}
Inserting Eq. (12) into Eq. (10) and right-multiplying by $e_{3}e_{4}$ produces:

\begin{equation}
e_{4}\partial^{4} \Xi_{\pm}  + e_{\mu}\partial^{\mu} \Xi_{\mp} = m \Xi_{\mp}Ee_{3}e_{4}  
\end{equation}

Now assume the ``cylinder'' condition Eq. (6) holds: $\Phi$, and therefore
also $\Xi_{\pm}$, does not depend explicitly on $x^{4}$. We can now drop the
first term in Eq. (13), which 
represents the coupling between $\Xi_{+}$
and $\Xi_{-}$. Thus, (writing out $e_{3}e_{4}E$ in full and
setting $\Xi_{\pm}=\Psi$),
we are left with two copies of the equation:

\begin{equation}
e_{\mu} \partial ^ {\mu}  \Psi = m \Psi e_{0} e_{1} e_{2} 
\end{equation}
which will be recognised as (3), the free-particle Dirac equation in Hestenes' form.

We note that Eq. (13) somewhat resembles the spin one-half wave equation of Dirac's
six-dimensional theory [18] mentioned above, in which the higher dimensions
are interpreted conformally.  In the Clifford approach to conformal geometry [38], idempotents
like those in Eq. (11) play an important role.  This suggests that the symmetries  of the
conformal group may underlie the seemingly rather arbitrary relationship Eq. (12)
required for our theory to have the correct limit.   

We have seen that Eq. (5) is equivalent to the standard free-particle
Dirac equation (3). We would now like to find a similar five-dimensional 
equation which is equivalent to the Dirac equation in the presence of an
external vector potential.

To accomplish this, we assume that the equation we seek has the form

\begin{equation}
e_{A} \partial^{A} \Phi = - m \Phi E  + qA\Phi\Gamma
\end{equation} 
where $q$ is a scalar charge, $A$ is a (five-dimensional) vector potential,
and $\Gamma$ is a blade to be determined. We see that $\Gamma$ must be
of even grade, since the grade of each term in the sum must be odd.
We will try to find a $\Gamma$  which, after applying 
the cylinder condition and setting $A^{4} = 0$, makes (15) identical
to the standard Hestenes-Dirac equation  for this case [36]:
\begin{equation}
e_{\mu} \partial ^ {\mu}  \Psi = m \Psi e_{0} e_{1} e_{2} + qA\Psi e_{1}e_{2}
\end{equation}

Using the rejection/projection operators and assuming the cylinder condition,
we obtain from (15)

\begin{equation}
e_{\mu} \partial ^ {\mu}  \Psi = m \Psi e_{3}e_{4}E + qA[(\Phi\Gamma)_{+} - (\Phi\Gamma)_{-}e_{3}e_{4} ]
\end{equation}
For this to agree with Eq. (16), we must have
\begin{equation}
\Phi_{+} - \Phi_{-}e_{3}e_{4} = (\Phi\Gamma)_{-} e_{1}e_{2}e_{3}e_{4} - (\Phi\Gamma)_{+}e_{1}e_{2}
\end{equation}
There are two possible solutions, depending on whether $\Gamma$ does or does not 
contain a factor of $e_{4}$, viz. 
$\Gamma = e_{1}e_{2}$
and
$\Gamma =  - e_{1}e_{2}e_{3}e_{4} = e_{0}E$

\section{Plane Wave Representation}

It may be seen by substitution that
Equation (5) has  plane wave solutions of  the form

\begin{equation}
\Phi = \phi(\cos(k^{A}x_{A}) + \Gamma \sin(k^{A}x_{A}))  
\end{equation}
where $\phi$ is a constant spinor amplitude and $\Gamma$ is any blade such that 
 $\Gamma\Gamma = -1$ and 
\begin{equation}
k^{A}e_{A}\Phi \Gamma = -m E \Phi  
\end{equation}

We want these waves to represent the free Dirac particle, i.e. for 
the $\Psi$ derived from Eq. (19) to  be the same (after the cylinder condition has been applied) 
as that for the ordinary Dirac plane wave [39]
\begin{equation}
\Psi = \psi(\cos(k^{\mu}x_{\mu}) + e_{1}e_{2} \sin(k^{\mu}x_{\mu}))  
\end{equation}
Using the rejection/projection operations on Eq. (19) and noting that by the cylinder condition
$k^{4}=0$, we find the following two criteria

\begin{equation}
\psi = \phi_{+} - \phi_{-}e_{3}e_{4}
\end{equation}
and
\begin{equation}
\psi e_{1}e_{2} = (\phi\Gamma)_{+} - (\phi\Gamma)_{-}e_{3}e_{4}
\end{equation}
that is:
\begin{equation}
(\phi_{+} - \phi_{-}e_{3}e_{4} )e_{1}e_{2} = (\phi\Gamma)_{+} - (\phi\Gamma)_{-}e_{3}e_{4}
\end{equation}

Now it is evident that $e_{4}$ can occur at most once in the spinor (that is, in the
\begin{em}even\end{em} grade blade) $\Gamma$. Let us first assume that $e_{4}$ is a
factor of $\Gamma$, so that there is an odd number of factors of the type $e_{\mu}$. 
Then by Eq. (9):

\begin{equation}
(\phi\Gamma)_{\pm} =  \phi_{\mp} \Gamma
\end{equation}
Inserting this into Eq. (15), we find that 

\begin{equation}
\Gamma =  - e_{1}e_{2}e_{3}e_{4} = e_{0}E
\end{equation}
while Eq. (20) becomes

\begin{equation}
k^{A}e_{A}\Phi  = m \Phi e_{0}  
\end{equation}

But this is not the only solution! We could also have assumed that $\Gamma$ 
is composed entirely of vectors orthogonal to $e_{4}$.  In that case, Eq. (25)
must be replaced by

\begin{equation}
(\phi\Gamma)_{\pm} =  \phi_{\pm} \Gamma
\end{equation}
and we see that

\begin{equation}
\Gamma = e_{1}e_{2}
\end{equation}
Similarly Eq. (26) becomes

\begin{equation}
k^{A}e_{A}\Phi  = - m \Phi e_{0}e_{3}e_{4}  
\end{equation}
If we insert a completely
general spinor $\Gamma$ into Eq. (15), we find that 
\begin{equation}
\Gamma_{+} - \Gamma_{-}e_{3}e_{4}   = e_{1}e_{2}
\end{equation}
Evidently, when $\Gamma_{-} = 0$, we are left with $\Gamma_{+} = e_{1}e_{2}$,
and when  $\Gamma_{+} = 0$, we are left with $\Gamma_{-} = e_{0}E$. We must 
therefore in general 
set  $\Gamma$ equal to a linear combination of  the two values given in
Eqs. (26) and (29).  Inserting such a superposition into Eq. (31) reveals that the
two scalar co\"{e}fficients add up to unity, and thus that we may think of them as
the squared sine and cosine of  some phase angle $\theta$:  
\begin{equation}
\Gamma=  e_{1}e_{2} ( \cos^{2} \theta -   e_{3}e_{4} \sin^{2} \theta)
\end{equation}
Such a 
$\Gamma$   satisfies Eq. (21) as expected. However,  it will be recalled 
that  $\Gamma\Gamma = -1 $.  This condition
and Eq. (32) can hold simultaneously only if $\theta$ is an integer 
multiple of $\pi/2$, that is to say,  if $\Gamma$ takes one of the 
two values specified in Eqs. (26) and (29).  

Thus, we have two separate classes of plane waves, \begin{em}both\end{em}
of which, thanks to the cylinder condition, correspond to the same 
ordinary Dirac plane waves. It will be noted that the values of 
$\Gamma$ are the same which arose in the treatment of the vector potential above.

\section{The Coulomb Potential}

Consider an electron moving in a spherically symmetric external potential 
\begin{equation}
qA=-\frac{\lambda}{|r|}e_{0}
\end{equation} 
where $\lambda$ is a scalar constant. Then, with the cylinder condition, Eq. (15) becomes:

\begin{equation}
e_{\mu} \partial^{\mu} \Phi = - m \Phi E  - \frac{\lambda}{|r|}e_{0}\Phi \Gamma 
\end{equation} 
The only specific properties of
$\Gamma$ we will need in this section are that $\Gamma\Gamma = -1$
and $\Gamma e_{0}= e_{0} \Gamma $.
Eq. (34) may be rewritten as:

\begin{equation}
\nabla  \Phi - e_{0} \partial_{t} \Phi = -m \Phi E + \frac{\lambda}{|r|}\zeta   \Phi e_{0}\Gamma
\end{equation}
where
\begin{equation}
\zeta   F\equiv e_{0}F e_{0}
\end{equation}
for any multivector $F$. Note that $\zeta   \zeta   F=F$.

As in the usual four-dimensional Dirac Coulomb problem, we will assume that
the energy $\varepsilon$ is related to the time-derivative of the state function.
In non-relativistic quantum mechanics, we know that:
\begin{equation}
\varepsilon \Psi  =\iota \partial_{t} \Psi   
\end{equation} 
(with $\hbar = 1$). In the Dirac theory as formulated by Hestenes
in the language of geometric algebra [32-36], the imaginary scalar $\iota$ is replaced
by the real bivector $e_{1}e_{2}$ (acting on $\Psi$ from 
the right); clearly, this squares to $-1$ as expected. 
There is no obvious prescription specifying uniquely what substitution to make in the
five-dimensional theory, but an obvious choice is $\Gamma$. This choice of course
coincides with the Dirac-Hestenes choice in the case that $\Gamma$ is given by $e_{1}e_{2}$,
but even when $\Gamma$ is given by $e_{0}E$ the proposed substitution will give the 
same final results as the conventional Dirac equation after the cylinder condition is 
applied. We therefore write:  
\begin{equation}
\varepsilon \Phi = \partial_{t} \Phi \Gamma 
\end{equation} 
Substituting this into Eq. (35) and right-multiplying by $\Gamma e_{0}$, we get
\begin{equation}
\eta   \nabla  \Phi - \varepsilon\zeta   \Phi = -m\eta   \Phi E + \frac{\lambda}{|r|}\zeta   \Phi 
\end{equation}
where
\begin{equation}
\eta  F\equiv F \Gamma e_{0} 
\end{equation}
for any multivector $F$. Note that $\eta  \eta  F=F$ and $\eta  \zeta    F = \zeta   \eta   F$.

To solve Eq. (40), we employ a geometric method analogous but not identical to that
used by Temple and Eddington [36, 40, 41] to solve the  conventional 
four-dimensional Coulomb problem. 

We make the assumption (justified by the cylinder condition and the requirement that the
final result reduce to Dirac's) that, like $\Psi$ in the four-dimensional case, $\Phi$ is an eigenfunction 
of the relativistic angular momentum operator with eigenvalue $\kappa$. Therefore [36]:
\begin{equation}
r \nabla \Phi = (r \bullet \nabla + 1 - \kappa \zeta   ) \Phi  
\end{equation}
Using the properties of the operators
$\zeta   $ and $\eta  $, we find:
\begin{equation}
\eta   r \bullet \nabla  \Phi + \eta  \Phi - \kappa \eta  \zeta    \Phi - \varepsilon r \zeta   \Phi = -E m r \eta   \Phi + \lambda e_{r} \zeta   \Phi 
\end{equation}
where $e_{r} \equiv r/|r|$.

To eliminate the second term, we change to a new variable  
\begin{equation}     
u \equiv |r| \Phi                                      
\end{equation} 
and introduce two new operators $S$ and $T$ by:
\begin{equation}
SF \equiv (\kappa + \lambda \eta  e_{r})\zeta   F 
\end{equation}
and
\begin{equation}
TF \equiv Ee_{r}(m-E\varepsilon\eta  \zeta   )F
\end{equation}
for any multivector $F$. These
greatly simplifies the appearance of Eq. (41), which becomes just: 
\begin{equation}
\partial_{r}u = \frac{1}{|r|}Su - Tu
\end{equation}

 We now expand $u$ in 
powers of $|r|$,and proceed  exactly as  in the Temple-Eddington method of solving
the four-dimensional Coulomb problem:
\begin{equation}
u =\sum_{p=0}^{\xi}  |r|^{p+q} e^{\beta |r|} C_{p} 
\end{equation} 
where $\xi$ is an integer greater than zero, 
$q$ and $\beta$ are scalars to be determined, and $C_{p}$ is some constant
spinor.  Equating coefficients we eventually obtain:
\begin{equation}
\sqrt{m^{2}-\varepsilon^{2}} (\xi+\sqrt{\kappa^{2}-\lambda^{2}}) = \varepsilon \lambda
\end{equation}
Eq. (48) is standard in four-dimensional Dirac theory, and is easily 
solved for $\varepsilon$ to give the usual relativistic energy spectrum
[42-44] first derived by Sommerfeld [45]:

\begin{equation}   \varepsilon= 
m/\sqrt{1+\frac{\lambda^{2}} {(\xi + \sqrt{\kappa^{2}-\lambda^{2}} )^{2}} }
    \end{equation}
Here $\kappa$ and $\xi$ are related to the non-relativistic quantum numbers $n$ and $j$ by
\begin{equation}
n=|\kappa| + \xi
\end{equation}
and 
\begin{equation}
j = |\kappa| - \frac{1}{2}
\end{equation}
The energy levels of a bound particle 
in a Coulomb potential are calculated from Eq. (49) and are
of course, identical to those obtained from the Dirac equation.
 Interpreting  $m$ in Eq. (49)
as the mass of the electron times $c^{2}$ and $\lambda$ as
the atomic number times the fine structure constant, we may evaluate Eq. (49) 
numerically for hydrogen-like atoms. Subtracting off the rest mass gives
the bound state energy levels, which for hydrogen are easily found to be:
$$
1s_{1/2} : \kappa=-1 : \xi=0 : \epsilon = -13.06 eV
$$
$$
2s_{1/2} : \kappa=-1 : \xi=1 : \epsilon = -3.402 eV
$$
$$
2p_{1/2} : \kappa=1 : \xi=1 : \epsilon = -3.402 eV
$$
$$
2p_{3/2} : \kappa=-2 : \xi=0 : \epsilon = -3.401 eV
$$
and so on as found in almost every textbook of relativistic quantum mechanics.
However,  we have obtained these  standard
results directly from the five-dimensional Eq. (15),
\begin{em}not\end{em} from the four-dimensional Dirac equation (16).

\section{Wave Equation without Cylinder Condition}

The method presented here can be viewed in two quite different ways: as a
technique for discovering solutions of the conventional Dirac equation, or
as a wave equation actually obeyed by fermions, and containing the Dirac 
equation as a special case. 

If one chooses to adopt the first viewpoint, the interest of our approach
lies in its use of the pseudoscalar operator $E$, which commutes with every 
multivector and thus reduces somewhat the awkwardness of 
standard four-dimensional Clifford algebra approach. The fifth dimension itself,
from this point of view, is simply part of the formal apparatus;
because of the cylinder condition, it plays no role in the
final results, which are identical to those of the conventional Dirac equation.
Thus, Eq. (15) can be regarded as an auxiliary equation, and solving it
as simply a technique for solving the Dirac equation, in which all the actual
physics resides. An analogy might be made to the use of complex vectors, the
imaginary parts of which are eventually set to zero, to simplify calculations
in electromagnetism.

However, the successes of  string theory and brane theory [e.g. 46] have made 
higher-dimensional approaches in which the additional co\"{o}rdinates are
physical rather than merely abstract increasingly popular, and thus it is
worthwhile to consider the second viewpoint as well:
the possibility that Eq. (15) \begin{em}without\end{em} the cylinder condition is the
correct one-body description of  a fermion.

The main difficulty with such a position is the discrepancy between the eight components
of a spinor in $C\ell(3,2)$ and the observed number of fermionic degrees of freedom.
This problem was already encountered in 1929 by Lanczos [23], whose two coupled
quaternionic wave equations we have seen to be closely related to Eq. (10).  
G\"{u}rsey [47] 
much later suggested that the Lanczos equations predict isospin doublets, an interpretation
strongly endorsed  by Gsponer and Hurni [24]. Each of the two four-component spinors
in Lanczos' theory, according to this explanation, represents a Dirac particle/antiparticle state
with isospin up or down. Clearly a similar interpretation could be applied to the two
halves of the single eight-component spinor in the present work: recall that  the projected
parts $\Phi_{+}$ and $\Phi_{-}$ reduce to Lanczos'  spinors when the cylinder condition is
applied.

Indeed, one would expect 
either a new class of particles  or a new quantum number
to arise in going from four  spinor components to eight, just as antiparticles
arise in going from two components to four and spin in going from one component to two. 
 If Eq. (15) rather than Eq. (16) is the correct description of a fermion, the degeneracy 
between $\Xi_{+}$
and $\Xi_{-}$ is lifted. Then it seems quite possible that each represents a different type of
particle (in the way that the ``large'' and ``small'' parts of the ordinary 
Dirac spinor represent particles and antiparticles), or else a previously unrecognised quantum 
state (in the way that the upper and
lower components of the Pauli spinor represent spin-up and spin-down states).      
Of course, with our cylinder condition enforced, the degeneracy becomes purely
formal, and both $\Xi_{\pm}$ are equivalent.

Similar considerations arise from the  fact that two quite
different approaches to the vector potential, given by the two choices
of $\Gamma$ (26) and (29), correspond to
the same four-dimensional result. This is unproblematic if the present
method is treated as merely a technique for solving the Dirac equation, but if 
the fifth dimension is an 
actual part of physical spacetime,  the ambiguity introduces  yet another degeneracy.
This is true even in the case of unbound particles, because of our freedom
to use either Eq. (26) or (29) in the phase.  Thus, dropping the cylinder condition,
or treating it as only an approximation,  would suggest   new physics
beyond the Dirac equation. 

One important prediction of the  five-dimensional wave equation (without the cylinder
condition) is the existence of a new scalar potential, which arises in any four-dimensional
region over which the $\Xi_{-}$ part of the wave function is constant. To show this,
we first maintain the upper sign in Eq. (13) and impose the constraint

\begin{equation}
\partial^{\mu}\Xi_{-} << 1  
\end{equation}
For non-zero mass, we obtain 

\begin{equation}
\Xi_{-}=\frac{1}{m} e_{4}\partial^{4} \Xi_{+} e_{0}e_{1}e_{2} 
\end{equation}
Inserting Eq. (53) now into Eq. (13) with the lower sign, we find that 

\begin{equation}
-\frac{1}{m}\partial^{4}\partial^{4} \Xi_{+} e_{0}e_{1}e_{2}  + e_{\mu}\partial^{\mu} \Xi_{+} = m \Xi_{+}e_{0}e_{1}e_{2}   
\end{equation}
The standard four-dimensional Hestenes-Dirac equation with a scalar potential $s$ is
given by:
\begin{equation}
-s \Psi e_{0}e_{1}e_{2}  + e_{\mu}\partial^{\mu} \Psi = m \Psi e_{0}e_{1}e_{2}   
\end{equation}
Comparing Eq. (54) to Eq. (55), we observe that the two are
identical if we replace $\Xi_{+}$ by $\Psi$ as we did in Section III
and identify $ms$ with the eigenvalue of the second partial 
derivative of $\Psi$: 

\begin{equation}
\partial^{4}\partial^{4} \Psi = ms \Psi.   
\end{equation} 
Thus  the dependence of $\Psi$ on the 
fifth dimension will manifest itself in four dimensions as a scalar potential.    
The scalar potential $s$ could perhaps be identified with the Higgs field, or with one of the scalar potentials
giving rise to inflationary processes in cosmology, Often the existence of 
such physically important scalars is  merely postulated, whereas in 
this approach, a 
scalar potential arises directly and necessarily from the wave equation itself!  

This procedure works only in that case of non-zero mass. For $m=0$, Eq. (13) 
with the restraint (52) becomes for the upper sign 

\begin{equation}
e_{4}\partial^{4}\Xi_{+}=0 
\end{equation}
i.e., for a massless particle, requiring $\Xi_{-}$ to be constant over a space-time region is the same as
requiring the cylinder condition to hold for $\Xi_{+}$ in the  region. 

The equation with the lower sign 
in the same case is
interesting for a different reason: it may be thought of 
as a four-dimensional Dirac equation with ``sources''. The ordinary four-dimensional Dirac-Hestenes
equation for a massless particle (i.e. Eq. (3) with the right hand side set to zero)
is formally identical to Maxwell's equation in empty space. Both 
equations (Maxwell's and Dirac's) may be written as: 

\begin{equation}
e_{\mu}\partial^{\mu} \Psi =0 
\end{equation}    
the only difference being that in the Dirac case $\Psi$ is a general even-grade multivector
of the form:
\begin{equation}
\Psi \equiv (\alpha + e_{0} e_{1} e_{2}e_{3} \beta) + \frac{1}{2} e_{\mu} \wedge e_{\nu} F^{\mu \nu}   
\end{equation}
where $\alpha$, $\beta$, and the six $F^{\mu \nu}$ are the components of the wave function,
whereas
in the Maxwell case $\Psi$ is restricted to grade two:

\begin{equation}
\Psi \equiv \frac{1}{2} e_{\mu} \wedge e_{\nu} F^{\mu \nu}  
\end{equation}
where $F^{\mu \nu}$ is the electromagnetic field tensor. However the Maxwell equation with 
source current $J$ given by: 

\begin{equation}
e_{\mu}\partial^{\mu} \Psi = -4 \pi J  
\end{equation}
has no counterpart in the conventional Dirac theory. 
On the other hand, we have from Eq. (13) with the lower sign and zero mass, writing $\Psi$ for $\Xi_{+}$:

\begin{equation}
 e_{\mu}\partial^{\mu} \Psi = - e_{4}\partial^{4} \Xi_{-}   
\end{equation}
so that formally Eqs. (61) and (62) are identical if 

\begin{equation}
J \equiv \frac {1}{4 \pi} e_{4}\partial^{4} \Xi_{-}  
\end{equation}
The constraint given by Eq. (52) guarantees that $J$ is a conserved quantity in
four dimensions. 

It may be recalled that $\Xi_{-}$ is the sum of a pseudoscalar term and various 
bivector terms all containing $e_{4}$. Consequently $e_{4}\partial^{4} \Xi_{-}$
(and therefore also $J$) is the sum of a pure space-time vector and a trivector
term \begin{em} not \end{em} containing $e_{4}$. Thus, just as the four-dimensional
massless Dirac equation differs from the vacuum Maxwell equation only by the 
presence of additional scalar and pseudoscalar terms in the field, so (for zero mass)
Eq. (13) with the lower sign differs further from the Maxwell equation with sources only by the 
presence of an additional trivector term in the source current. It seems reasonable to suppose that 
in both cases these differences may be related to the contrast between the integer spin of the photon  
and the half-integer spin of the Dirac particle.

In the Maxwell case, the source currents represent an external charge distribution.
Eq. (62), however, indicates that the source currents for Eq. (61) arise from the
$\Xi_{-}$ part of the wave function itself. This is somewhat reminiscent of de Broglie's
``double solution'' approach to the Schr\"{o}dinger equation [48], which has enjoyed
a resurgence of interest in the last decade because of its relationship to Bohmian
quantum mechanics [49]. Perhaps Eq. (61) provides a link between the solitons of
de Broglie's theory and the solitons in the five-dimensional induced matter theory 
of Wesson and Ponce de Leon [9, 13, 14].

\section{Conclusion}
 
We have shown that the simple five-dimensional wave equation Eq. (5) 
is a powerful tool for the study of fermions. When the cylinder condition is enforced,
this equation is exactly equivalent to the conventional Dirac equation, but is
in some respects more tractable; it thus provides a useful new technique for doing
relativistic quantum mechanics. The two solutions $\Xi_{+}$
and $\Xi_{-}$ of the projected five-dimensional equation are  degenerate, and either one 
may be identified with the conventional Dirac wave function. 

In the case of a free particle, we have discovered two families of plane-waves given by
Eq. (19) and
differing only in the two possible choices of $\Gamma$, Eqs. (26) and (29). 
As long as the cylinder condition holds, there is no way to distinguish these
two families, and either one (but not both at once) may be taken as a representation
of the free particle. Likewise, in the case of  a bound particle, we found the same two  choices of
$\Gamma$ satisfy Eq. (15). Either selection, with
the cylinder condition,   splits Eq. (15) into two identical copies of the Dirac equation. 
 Thus, solving (15) is the same as solving the 
Dirac equation, as long as the cylinder condition is in force, and we may choose
that value of $\Gamma$ which is most convenient for a given potential.

When the cylinder condition is not imposed, i.e. when the dependence on the
fifth dimension is retained, the number of spinor degrees of freedom
is doubled, suggesting the
possible existence of a new quantum number. In the limiting case where 
four of the spinor components are held constant and those remaining are 
identified with the four-dimensional wave function of a 
\begin{em}massive\end{em} fermion, Eq. (5) predicts the 
existence of a scalar field which (per unit mass) is simply the eigenvalue of
the operator $\partial^{4}\partial^{4}$. If the particle is instead assumed
to be massless,
Eq. (5) is formally identical to Maxwell's equation with sources; such
sources are absent in conventional fermionic quantum mechanics. Thus,
in both cases, new physical effects arise from the relaxation of the 
cylinder condition.

\newpage

\section{References}

\begin{enumerate}
\item P. A. M. Dirac, ``The Electron Wave Equation in de-Sitter Space.''
\begin{em}Ann. Math.\end{em} \begin{bf}36\end{bf},  657 (1935). 
\item A. Macias and H. Dehnen, ``Dirac Field in the Five-Dimensional Kaluza-Klein Theory.''
\begin{em} Class. Quantum Grav.\end{em} \begin{bf} 8\end{bf}, 203 (1991).
\item D. S. Staudte, ``An Eight-Component Relativistic Wave Equation for
Spin One-Half Particles.''
\begin{em}J. Phys.\end{em} \begin{bf}A 29 \end{bf},  169 (1996).  
\item J. Kocinski, ``A Five-Dimensional Form of the Dirac Equation.''
\begin{em}J. Phys.\end{em} \begin{bf}A 32\end{bf}, 457 (1999). 
\item J. B. Almeida, ``Choice of the Best Geometry to Explain Physics.'' 
ArXiv: physics/0510179 October 19, 2005.
\item J. B. Almeida, ``A Geometric Algebra Approach to the Hydrogen Atom.'' 
ArXiv: physics/0602116 February 16, 2006.
\item D. Bailin and A. Love, ``Kaluza-Klein Theories.''
\begin{em}Rep. Prog. Phys.\end{em} \begin{bf}50\end{bf},  1087 (1987).
\item J. Kocinski and M. Wierzbicki, ``The Schwarzschild Solution in a Kaluza-Klein Theory with Two Times.''
in \begin{em}Relativity, Gravitation, Cosmology,\end{em} edited by V. Dvoeglazov and
A. Espinoza Garrido (Nova Science, New York, 2004). 
\item P. S. Wesson, ``Wave Mechanics and General Relativity: A Rapprochement.''
\begin{em} Gen. Rel. Grav.\end{em} \begin{bf} 38\end{bf}, 937 (2006).
\item P. S. Wesson, ``The 4D Klein-Gordon, Dirac and Quantization Equations from 5D Null Paths.''
\begin{em} Gen. Rel. Grav.\end{em} \begin{bf} 35\end{bf}, 111 (2003).
\item P. S. Wesson, ``Five-Dimensional Relativity and Two Times.''
\begin{em}Phys. Lett.\end{em} B \begin{bf}538\end{bf}, 159  (2002).
\item A. Billyard and P. S. Wesson, ``A Class of Exact Solutions in 5D
Gravity and its Physical Properties.''\begin{em}Phys. Rev.\end{em} \begin{bf}D 53\end{bf}, 
731 (1996).
\item J. Ponce de Leon, ``Equivalence Between Space-Time-Matter and Brane-World Theories.''
\begin{em}Mod. Phys. Lett.\end{em} \begin{bf}A 16\end{bf},  2291 (2001).
\item J. Ponce de Leon, ``Kaluza-Klein Solitons Reexamined.''
ArXiv: gr-qc/0611082 November 15, 2006.
\item I. Bars and C. Kounnas, ``String and Particle with Two Times.''
\begin{em}Phys. Rev.\end{em} \begin{bf}D 56\end{bf},  3664 (1997).
\item I. Bars and C. Deliduman, ``High Spin Gauge Fields and Two-Time Physics.''
\begin{em}Phys. Rev.\end{em} \begin{bf}D 64\end{bf},  045004 (2001).
\item I. Bars, ``Survey of Two-Time Physics.''
\begin{em}Class. Quant. Grav.\end{em} \begin{bf}18\end{bf},  3113 (2001).
\item P. A. M. Dirac, ``Wave Equation in a Conformal Space.''
\begin{em}Ann. Math.\end{em} \begin{bf}37\end{bf},  429 (1936). 
\item H. A. Kastrup, ``Gauge Properties of Minkowski Space.''
\begin{em}Phys. Rev.\end{em} \begin{bf}150\end{bf},  1183 (1966).
\item G. Mack and A. Salam, ``Finite-Component Field Representation of the 
Conformal Group.''
\begin{em}Ann. Physics\end{em}  \begin{bf}53\end{bf},  174 (1969).
\item C. R. Preitschopf and M. A. Vasilev, 
``Conformal Field Theory in Conformal Space.''
\begin{em}Nucl. Phys.\end{em} \begin{bf}B 549\end{bf},  450 (1999).
\item W. P. Joyce, ``Dirac Theory in Spacetime Algebra.''
\begin{em}J. Phys. \end{em} \begin{bf}A 34\end{bf},  1991 (2001).
\item C. Lanczos, ``Die tensoranalytischen Beziehungen der Diracschen Gleichung.''
\begin{em}ZS. f. Phys.\end{em} \begin{bf}57\end{bf},  447 (1929). 
\item A. Gsponer and J.-P. Hurni, ``Lanczos-Einstein-Petiau: From
Dirac's Equation to Nonlinear Wave Mechanics,'' in 
\begin{em}Cornelius Lanczos: Collected Published Papers with Commentaries, \end{em}
edited by W. R. Davis, 
(North Carolina State University 
Press, Raleigh, 1998), Volume III, Part 2, pp. 1248-1277. 
\item W. K. Clifford, \begin{em}Collected Mathematical Papers\end{em} 
(Macmillan, London, 1882).
\item C. Doran and A. Lasenby, \begin{em}Geometric Algebra for Physicists\end{em}
(Cambridge University Press, 2003). 
\item E. Bayro Corrochano and G. Sobczyk, editors,
\begin{em}Geometric Algebra with Applications in Science and Engineering\end{em}
(Birkh\"{a}user, Boston, 2001).
\item P. Lounesto, \begin{em}Clifford Algebras and Spinors\end{em}
(Cambridge University Press, 1997). 
\item D. Hestenes and G. Sobczyk, \begin{em}Clifford Algebra to Geometric Calculus\end{em}
(Reidel, Dordrecht, 1984).
\item D. Hestenes, \begin{em}Space-Time Algebra\end{em}
(Gordon and Breach, New York, 1966).
\item S. Weinberg, \begin{em}Quantum Theory of Fields\end{em}
(Cambridge University Press, 1995), Volume I, p. 9.
\item D. Hestenes, ``Local Observables in the Dirac Theory.'' 
\begin{em}J. Math. Phys.\end{em} \begin{bf}14\end{bf},  893 (1973). 
\item D. Hestenes, ``The Zitterbewegung Interpretation of Quantum Mechanics.''
\begin{em}Found. Phys.\end{em} \begin{bf}20\end{bf},  1213 (1990) .
\item P. Lounesto, ``Clifford Algebras and Hestenes Spinors.''
\begin{em}Found. Phys.\end{em} \begin{bf}23\end{bf},  1203 (1993).
\item W. P. Joyce and J. G. Martin, ``Equivalence of Dirac Formulations.''
\begin{em}J. Phys. \end{em} \begin{bf}A  35\end{bf},  4729 (2002).
\item Reference [26], Chapter 8.
\item C. Doran, A. Lasenby, S. Gull, S. Somaroo, and A. Challinor, 
``Spacetime Algebra and Electron Physics,'' in 
\begin{em}Advances in Imaging and Electron Physics, \end{em} 
edited by P. W. Hawkes,  (Academic Press,  New York, 1996), 
Vol. 95, p. 271.
\item C. Doran, A. Lasenby, and J. Lasenby, 
``Conformal Geometry, Euclidean Space and Geometric Algebra,'' in 
\begin{em}Uncertainty in Geometric Computations, \end{em} edited by J. Winkler,  
(Dordrecht: Kluwer, 2002), p. 41.
\item D. Hestenes, 
``Zitterbewegung in Radiative Processes,'' in 
\begin{em}The Electron ,\end{em} edited by 
D. Hestenes and 
A. Weingartshofer, (Kluwer, Dordrecht, 1991),  p. 21.
\item G. Temple, ``The Operational Wave Equation and the Energy Levels of
the Hydrogen Atom.'' \begin{em}Proc. Roy. Soc.\end{em} 
\begin{bf}A 127\end{bf},  349 (1930).
\item A. S. Eddington, \begin{em}Relativity Theory of Protons and Electrons,\end{em}
(Cambridge University Press, 1936), Chapter 9.
\item P. A. M. Dirac, ``Quantum Mechanics and a Preliminary Investigation of
the Hydrogen Atom.'' \begin{em}Proc. Roy. Soc.\end{em} \begin{bf}A 110\end{bf}, 
561 (1926).
\item W. Greiner, \begin{em}Relativistic Quantum Mechanics,\end{em} 
(Springer Verlag, Heidelberg, 1990), p. 182.
\item P. Strange, \begin{em}Relativistic Quantum Mechanics,\end{em} 
(Cambridge University Press, 1998), p. 238.
\item A. Sommerfeld, ``Zur Quantentheorie der 
Spektrallinien.'' \begin{em}Ann. Physik\end{em} \begin{bf}51\end{bf}, 
1 and 125 (1916). 
\item R. Maartens, ``Brane-World Gravity.''
\begin{em}Living Rev. Relativity\end{em} \begin{bf}7\end{bf},  7 (2004).
\item F. G\"{u}rsey, ``On the Symmetries of Strong and Weak Interactions.''
\begin{em}Nuov. Cim.\end{em} \begin{bf}16\end{bf},  230 (1960).
\item L. V. de Broglie, \begin{em}Une tentative d'interpr\'{e}tation causale et non lin\'{e}aire 
de la m\'{e}canique ondulatoire,\end{em} 
(Gauthier-Villars, Paris, 1956).
\item A. Abbondandolo and V. Benci, ``Solitary Waves and Bohmian Mechanics.''
\begin{em}Proc. Nat. Acad. Sci. (USA)\end{em} \begin{bf}99\end{bf},  15257 (2002).

\end{enumerate}
\end{document}